\documentclass[12pt,a4paper]{article} 
\usepackage{bbm}
\usepackage{amsgen}
\usepackage{amsfonts}
\usepackage{amssymb}
\usepackage{amsbsy}
\usepackage{fullpage}
\usepackage{epsfig}

\def\figsubcap#1{\par\noindent\centering\footnotesize(#1)}

\def\N{\mathbbm{N}}

\renewcommand{\title}[1]{\topsep=0pt\begin{center}\Large\bf#1\end{center}\vspace{12pt}} 
\renewcommand{\author}[1]{\topsep=0pt\begin{center}\large\rm#1\end{center}} 
\newcommand{\address}[1]{\topsep=0pt\begin{center}\footnotesize\it#1\end{center}\vspace{12pt}} 
\renewcommand{\date}[1]{\topsep=0pt\begin{center}\small#1\end{center}\vspace{12pt}} 

\begin{document} 

\title{Reheating induced by competing decay modes} 
\author{T. Charters${}^{\dag}$, A. Nunes${}^{\ddag}$, J. P. Mimoso${}^{\S}$} 

\address{${}^{\dag}{}$ 
Departamento de Engenharia Mec\^anica/\'Area  Cient\' \i fica  de Matem\'atica\\ 
Instituto Superior de Engenharia de Lisboa\\ Rua Conselheiro 
Em\'{\i}dio Navarro, 1, P-1949-014 Lisbon, Portugal\\ 
Centro de F\'{\i}sica Te\'{o}rica e Computacional 
da Universidade de Lisboa \\ 
Avenida Professor Gama Pinto 2, P-1649-003 Lisbon, Portugal} 

\address{${}^{\ddag}$ ${}^{\S}$
Departamento de F\'{\i}sica, Faculdade de Ci\^encias da Universidade de Lisboa \\ 
Centro de F\'{\i}sica Te\'{o}rica e Computacional 
da Universidade de Lisboa \\ 
Avenida Professor Gama Pinto 2, P-1649-003 Lisbon, Portugal 
} 

\address{${}^\dag$tca@cii.fc.ul.pt, ${}^\ddag$anunes@ptmat.fc.ul.pt, 
${}^\S$jpmimoso@cii.fc.ul.pt} 

\date{\today} 

\abstract{We address the problem of studying the decay of the inflaton field 
$\phi$  to another scalar field  $\chi$
through parametric  resonance  in the case of a coupling that
involves several decay modes.
This amounts to the presence 
of extra  harmonic terms in the  perturbation of the $\chi$  field dynamics. 
For the case of two frequencies we compute the geometry of the resonance 
regions, which  is significantly  altered due to the presence of
non-cuspidal resonance regions associated to higher harmonics and to
the emergence  of instability
`pockets'.  We discuss the effect of this change in the efficiency 
of the energy transfer process for the  simplest
case of a coupling given by a combination of the two 
interaction terms of homogeneous degree
usually considered in the literature.
We  find  that  the presence of 
higher harmonics has  limited cosmological implications. } 

\section{Introduction} 


The success of the inflationary paradigm depends to a great extent on the
corresponding success of the reheating stage that takes place  after  
inflation \cite{Shtanov:95,Kofman:1997b} through which  all elementary particles
that exist in the universe were created.  
During inflation the universe expands exponentially, its matter content is 
diluted and the temperature decreases as the inverse of the exponential of the
number of e-folds $\exp{N_e}$. 
Unless some different mechanism is considered for the inflaton decay (as in warm inflation
 \cite{Berera 95a}, \cite{Mimoso:2005bv}) 
there has to be some process 
to raise the temperature to the levels required for the nucleosynthesis of the light 
elements to take place according to the standard thermal history of the big-bang universe \cite{Liddle+Lyth:2000}.

In most post-inflationary scenarios reheating occurs due to particle production by an 
oscillating scalar field $\phi$.  In the  simplest models this field is  the 
inflaton  field that drives  inflation. 
After inflation the scalar field $\phi$ oscillates near a minimum of its potential and 
this triggers a sequence of processes that produces elementary particles 
and eventually restores the temperature 
 \cite{Kofman:1994rk,Boyanovsky+1:95,Boyanovsky+2:95,Yoshimura:1995gc,Fujisaki:1995dy,Chung:1998rq,Berges:2002cz}. 

Since the beginning of the 90's a considerable effort has been devoted to model the
reheating process \cite{Dolgov:1989us,Traschen:90,Kofman:1994rk} (a general
account can be found in  \cite{Kofman:1997b}). In most models, the first stage
of this complicated sequence involves the excitation by parametric resonance of
a second scalar field, here denoted $\chi$,  giving  rise to an exponential
increase in the  
number of $\chi$ boson particles \cite{Kaiser:1997hg,Greene:1997fu} (for a comprehensive
review see also  \cite{Bassett:2005xm}). 

Despite the many contributions regarding the preheating/reheating mechanism
itself
\cite{Anderson:1996kr,Son:1996uv,Kasuya:1997ha,Khlebnikov:1998sz,Micha:2002ey,Micha:2003ws,Podolsky:2005bw,Charters:2005eg,Felder:2006cc,GarciaBellido:2007dg}
and its observational implications
\cite{Finelli:1998bu,Tsujikawa:2002nf,Jokinen:2005by,Sa:2007pc}, some general
questions  remain to be completely answered. In  particular, how does the
resonant energy transfer depend on the  
coupling between $\phi$\ and $\chi$ and on the inflaton asymptotic dynamics? 
The latter issue  has been studied in  depth for a large variety  of polynomial 
potentials
\cite{Greene:1997fu,Felder:2000hj,Taruya:1997iv,Desroche:2005yt}. The former
issue is less well studied and has been the subject of some recent work
\cite{Dufaux:2006ee,BasteroGil:2007mm,ArmendarizPicon:2007iv}. 

In the simplest of the pre-heating scenarios the inflaton couples to the $\chi$ 
field through interaction terms of the form $h \phi \chi^2$ or $g^2 \phi^2 
\chi^2$, that correspond to two different decay modes of the 
scalar field $\phi$ into another boson \cite{Shtanov:95,Kofman:1997b}. These two
coupling terms give rise to the same qualitative effects, and are   
considered as alternative models. Indeed, in both cases the equation for 
the scalar field $\chi$ can be reduced to a Mathieu equation, and thus the 
parametric resonance follows similar patterns (even though the numerical values 
of the model's outcome may be slightly different). 
There are two regimes, a 
broad resonance regime, in which the amplitude of the periodic perturbation of 
the $\chi$-field frequency is of the same order as or larger than the frequency of 
the $\phi$ scalar field, and a narrow resonance 
regime where the amplitude of the perturbation is small. The broad resonance region
of parameter space includes, for sufficiently large values of
the perturbation, the tachionic resonance regime which  has been shown in
\cite{Dufaux:2006ee} to be extremely effective in transfering most of the energy
of the inflation to the $\chi$ field.

The main feature that emerges from these studies of pre-heating is that the 
broad and tachyonic resonance regimes gives the predominant contribution to the
$\chi $ field energy density. However, there is always the possibility of a
contribution to the total particle production in the narrow resonance regime
when the coupling parameters
characterizing the interaction between $\phi$ and $\chi$ are small and/or in the
decay of residual inflaton oscillations. In  the particular case of 
parametric resonance modelled by a Mathieu equation, this contribution is indeed small.
The growth of the modes of $\chi$ is exponential in the resonance bands or
tongues in parameter space, and the first resonance band is the only band wide
enough to give rise to significant $\chi$ excitations. 
However, this need not be so when there are more frequencies of 
excitation of the $\chi$ field. Here we show that in this case resonance
is governed by a general Hill equation  and that 
other resonances beyond the first may contribute to the amplification of the $\chi$ modes.
Therefore, there is the 
possibility that these higher frequency excitations contribute significantly 
to the overall creation of $\chi$ particles. This analysis is the subject of the 
present work, in the case when the $\chi$ field is parametrically forced by two
frequencies in a $1:2$ ratio. For this case, we show that the contribution for
the $\chi$ field energy density of the higher harmonic is a small fraction of
that of the fundamental frequency. 

The outline of the paper is as follows. In Section 2 we review the 
parametric frequency pre-heating mechanism and the method to compute
the particle production rate 
in the general framework of Hill's equation. 
In Section 3 we apply this method to compare the reheating efficiency of 
two different couplings of the inflaton field to the $\chi $ field.
In Section 4 we sum up the conclusions of this analysis.

\section{Particle production by parametric resonance} 
\label{particle.ppr} 
  
We start by reviewing  some properties of the parametric resonance mechanism 
for a general periodic perturbation of the frequency of the oscillator, which corresponds to  Hill's equation (\ref{eq:chi.hill}). 

Reheating models in inflationary universes start by considering that, at the end of inflation, the inflaton $\phi$ is in a coherent oscillatory state described by a space-independent expectation value, governed by the equation of motion for the inflaton  
\begin{eqnarray} 
 \label{eq:phi.frw} 
 \ddot \phi + 3 H \dot \phi + a^{-2}\nabla^2\phi +V'(\phi)=0, 
\end{eqnarray} 
where $H$ is the Hubble parameter of a Friedman-Robertson-Walker metric with 
scale factor $a(t)$, and $V'(\phi)$ is the derivative of the inflaton's potential $V(\phi)$ 
with respect to $\phi$. It is assumed that $V(\phi)$ has a vanishing minimum for the oscillations to take place, and also that the inflaton couples to another scalar field $\chi$ which is then periodically perturbed by the inflaton. The equation of motion of $\chi$ is given by 
\begin{eqnarray} 
 \label{eq:chi} 
 \ddot \chi + 3 H \dot \chi + a^{-2}\nabla^2\chi 
 +U'(\chi)+\frac{\partial}{\partial \chi}V_{int}(\phi,\chi)=0,
\end{eqnarray} 
where $V_{int}=V_{int}(\phi,\chi)$ is the interaction potential between $\chi$
and $\phi$, and where $U(\chi)$ is a potential that gives mass to $\chi$. At the
onset of the process of energy  
tranfer from $\phi$,  $\chi$ is assumed to be at the vanishing minimum of this
potential. The oscillations of $\phi$ around the minimum of its potential are
faster than the expansion rate of the universe, so it is meaningful in a
relatively short time scale to 
work within the simplifying assumption that the actual spacetime can be approximated 
with a Minkowski metric.
For reasonable choices of a single interaction potential $V_{int}$, namely, cubic interactions 
of   the  form   $g\phi\chi^2$,  or quartic interactions $h\phi^2 \chi^2$, the  resulting $\chi$ equation (\ref{eq:chi}) is that of an oscillator with a harmonically
 perturbed frequency. This yields a Mathieu type equation and provides a
 well-known mechanism for the parametric resonance of $\chi$ and for the
 exponential amplification of its particle number
 \cite{Bassett:2005xm,Charters:2005eg}.  
 
For more general coupling terms, in particular non-homogeneous couplings involving different powers
of $\phi $, the perturbation equation for $\chi$ 
is parametrically forced by a  periodic function (for the quasi-periodic case see for instance 
 \cite{Basset:98})  and the equations of  motion for the $\chi $ modes are of the form
(more details in Section \ref{pocket.inst})
\begin{eqnarray} 
 \label{eq:chi.hill} 
 \ddot\chi_k + (\omega_k^2 + \epsilon F(t))\chi_k = 0,
\label{hill}
\end{eqnarray} 
where $F(t)$ is a periodic function with period $2\pi$, 
and we may set $F(t)=F(-t)$, and $\int_0^{2\pi} F(t)dt =0$, 
and $\epsilon$ a small positive parameter related with the amplitude of the inflaton
oscillations.

For the general Hill equation (\ref{hill}), Floquet's theorem states that the solutions are of the form
\begin{equation}
\chi_k(t)=e^{\mu_k t} \chi_k^{(0)}(t)  ,
\end{equation}
where $\chi_k^{(0)}(t)$ is a periodic function with the same period as $F(t)$ and $\mu _k$
is one of the two characteristic exponents, which are both real or complex conjugate.
 
Clearly, the lines ${\rm Re} (\mu_k)=0$ divide the $(\omega_k, \epsilon)$ parameter plane
in unstable regions and stable ones, defining instability bands which become narrow
close to the $\epsilon =0$ axis, producing what is often called a structure of tongues  \cite{Arnold}. For a perturbation of period $2\pi$ these tongues end  on the 
$\epsilon =0$ axis at the points $\omega_k$ that satisfy the parametric resonance condition
$\omega_k = n/2$ (see Figure \ref{fig1}-a). These modes are amplified by an arbitrarily small parametric forcing of period $2 \pi$. 

To determine the characteristic exponent of the solutions of equation 
(\ref{eq:chi.hill}) we follow Hill's method of solution.
Given that the function $F$ is periodic we write it as 
\begin{eqnarray} 
 \label{eq:F} 
 F(t)=\frac{1}{2}\sum_{k=-s}^s c_k e^{ikt},
\label{Ff} 
\end{eqnarray} 
where $c_0=0$ (due to the parity of $F$) and $s \in \N \bigcup \{\infty \}$. 
This (finite or infinite) Fourier series expansion of $F(t)$, (\ref{eq:F}), 
together with Floquet's theorem suggest looking for a solution of 
(\ref{eq:chi.hill}) of the form 
\begin{eqnarray} 
\label{fourieru} 
\chi_k(t)=e^{\mu_k t}\sum_{n=-\infty}^{+\infty} b_n e^{int}. 
\end{eqnarray} 
Inserting this expression and  equation (\ref{eq:F}) in (\ref{eq:chi.hill}), and 
equating the coefficients of $e^{(\mu_k+in)t}$, we derive a homogeneous system with infinitely many 
linear equations 
\begin{eqnarray} 
\label{bneq} 
(\mu_k + in)^2b_n +\frac{\epsilon}{2} \sum_{m=-\infty}^{+\infty}c_mb_{n-m}=0,\ 
 n=\ldots,-2,-1,0,1,2,\ldots 
\end{eqnarray} 
Eliminating the coefficients $b_n$ in (\ref{bneq}), a non trivial solution exists if
the characteristic exponent satisfies
an infinite 
determinantal equation (called Hill's determinantal equation), 
\begin{eqnarray} 
\label{hilldet} 
\Delta(\epsilon,\mu_k) = \vert B_{rs}\vert = 0 
\end{eqnarray} 
where the elements $B_{rs}$  of
$\Delta(\epsilon,\mu_k)$ are given by
\begin{eqnarray} 
\label{eq:Brs} 
B_{rs}=\cases{1 & if $r=s$\cr \displaystyle 
\frac{\epsilon c_{r-s}/2}{(\mu+ir)^2+\omega_k} & if $r\ne s$}. 
\end{eqnarray} 
Here an infinite determinant $D=\vert B_{mn}\vert , 
(m,n=-\infty,\ldots,+\infty)$ is defined as the limit of $D_m = {\rm 
det}(B_{ij}) (i,j=-m,\ldots,m)$ as $m\to+\infty$, if it exists.

Equation 
(\ref{hilldet}) can be reduced to the simpler form  
of a transcendental equation in $\mu_k$  \cite{HillsEqBook}
\begin{eqnarray} 
\label{simplehilldet} 
\sin^2(i \pi\mu_k)=\Delta(\epsilon, 0)\sin^2(\pi\omega_k) . 
\end{eqnarray} 
This determines a characteristic exponent $\mu_k$, which in turn 
determines the $b_n$ coefficients of (\ref{bneq}), and hence, a 
formal solution (\ref{fourieru}) of Hill's equation. The real part of $\mu_k$ determines the growth factor of the solutions (\ref{fourieru}). 

In the narrow resonance regime the phenomenon of parametrically resonant
excitations, where $\epsilon \ll 1$, it is possible to derive 
an approximate expression in closed form for $\mu_k$ as a 
function of $\epsilon$. First, notice that $\Delta(0,0)=1$. Second, we see from (\ref{eq:Brs}) that the values of $B_{rs}$ depend  linearly on $\epsilon$, and
that
$\Delta(\epsilon,0)=1+O(\epsilon^2)$. This means  $\Delta(\epsilon,0)\simeq 1 + 
\alpha  \epsilon^2$, where $\alpha$ depends on 
$\mathbf{c}=(c_{-s},c_{-s+1},\ldots,c_{-1},c_{1},\ldots,c_{s-1},c_s)$ and on $\omega_k$. 

Inverting  (\ref{simplehilldet})
using $\arcsin z=\ln\left(i z +\sqrt{1-z^2}\right)$ and expanding in powers of 
$\epsilon$, yields for the growth factor
\begin{eqnarray} 
 \label{eq:mue} 
 {\rm Re}(\mu_k)\simeq \epsilon \alpha(\mathbf{c},\omega_k)  
 \left\vert\sin\left(\pi\omega_k\right)\right\vert. 
\end{eqnarray} 
This expression will be used later to determine numerically the resonant tongues
for a particular function $F(t)$. 

For periodic perturbations with finite Fourier expansions (\ref{Ff}), 
$s \in \N$, 
it can be shown that the asymptotic form of the width $L_n$ of the  $n$-th interval of instability is 
given by  \cite{HillsEqBook} 
\begin{eqnarray} 
 \label{eq:Ln} 
 L_n&=&\frac{8s^2}{[(p-1)!]}\left(\frac{\vert c_s\epsilon\vert}{8s^2 4^p}\right)^p+ 
 O(\epsilon^{p+1}),\quad n=sp\\ 
 L_n&=& a_n\frac{\vert\epsilon\vert^p}{4^p}+O(\epsilon^{p+1}),\quad s(p-1)<n< sp, 
\label{widths}
\end{eqnarray} 
where $p$ is the integer defined in these equations and where $a_n$ is a constant independent 
of $\epsilon$. One of the remarkable consequences of these formulae is that 
the widths of all the resonant bands
up to the  $s$-th band depend linearly on the perturbation amplitude, while the
resonances of order higher than $s$ are associated with thinner instability regions. 

This has direct impact in the computation of the energy production by parametric resonance.
For small $\epsilon $ the main contributions to the 
energy density of the $\chi$ field 
\begin{eqnarray} 
 \rho_\chi(t)&=& \frac{1}{2}\int d^3 \mathbf{k}\left[ 
   \omega_k^2 \vert \chi_k\vert^2+\vert \dot\chi_k\vert^2\right]\\ 
 &\simeq&\int d^3\mathbf{k} \omega_k^2 \vert \chi_k(t)\vert^2
\end{eqnarray} 
come from the bands with widths depending linearly on the 
perturbation amplitude $\epsilon$, since the other have 
relatively negligible widths due their nonlinear cuspidal form. 

This yields 
\begin{eqnarray} 
 \label{eq:Echi} 
 \rho_\chi(t) 
 &\simeq& 4\pi \left.\sum_{m=1}^s k^2\omega_k^2 L_m 
   \vert\chi_k(0)\vert^2 e^{\mu_k t}\right\vert_{\omega_k\simeq m/2}   ,
\end{eqnarray} 
where the value of $\chi_k(0)$ is evaluated at the center of the resonances 
bands. 

If all the modes start out with an amplitude $\vert\chi_k(0)\vert$ then, 
by the virial theorem, or assuming that there is one 
$\chi$ particle on each mode,   
\begin{eqnarray} 
 \label{eq:chi02} 
 \vert\chi_k(0)\vert^2=\frac{1}{\omega_k}, 
\end{eqnarray} 
and  so
\begin{eqnarray} 
 \label{eq:rhochi} 
  \rho_\chi(t)\simeq 4\pi \left.\sum_{i=1}^s k^2\omega_k L_i 
 e^{\mu_k t}\right\vert_{\omega_k\simeq i/2}. 
\end{eqnarray} 

The growth of the $\chi$ field modes persists in an expanding universe, 
if the time scale for the resonance is much shorter than the expanding time scale \cite{Traschen:90}. 
If we change the field to 
\begin{eqnarray} 
 \psi_k=a\chi_k, 
\end{eqnarray} 
and introduce conformal time $\eta$, defined by $d\eta=m a^{-1}dt$, the equation 
(\ref{eq:chi}) for $\psi_k$ becomes
\begin{eqnarray} 
 \label{eq:psi} 
 \psi_k''+\left(\frac{k^2}{m^2} + \epsilon 
       F(\eta)a^2-\frac{a''}{m^2 a}\right)\psi_k=0  ,
\end{eqnarray} 
where the prime denotes the derivative with respect to $\eta$. 

Assuming that $m\gg H$, we can treat the expansion of the universe adiabatically, 
and thus, at any given time, its effect is a shift in the oscillatory frequency 
\begin{eqnarray} 
 \label{eq:ktok} 
 k^2\to k^2-\frac{a''}{a}, 
\end{eqnarray} 
and also an adiabatic increase of the amplitude of the driving force. 
In the expanding universe the parametric resonance analysis is only applicable if 
several conditions are satisfied: 
\emph{(i)} $\epsilon/m^2\ll 1$ (perturbative regime) , 
\emph{(ii)}  the expansion of the universe can be neglected, that is $H\ll m$, 
and 
\emph{(iii)} we also need the time scale on which the instable solution grows 
to be smaller than $H$, so that expansion is unimportant, $H/m\ll \epsilon$, and 
finally, 
\emph{(iv)} that the frequency does not redshift out of the resonance band in a time 
 interval shorter than the amplification period $m/(2\mu_k)$. This latter 
 condition, for the $i$-band, can be written in the form 
\begin{eqnarray} 
\label{eq:cond_omega} 
 \left\vert\frac{m}{2\mu_k} \frac{d}{d\eta} \omega_k\right\vert < L_i, 
\end{eqnarray} 
where $\mu_k$ and $\omega_k$ are evaluated at the center of the resonance $ith$-band. 

The regime where these conditions can be easily satisfied is the period after 
inflation where naturally one has $H/m\ll 1$ and  where the equation of motion, 
in conformal time, for  $\bar\phi =a \phi$, is given by 
\begin{eqnarray} 
 \label{eq:barphi} 
 \bar \phi'' +\bar \phi \left(1-\delta\right)=0, 
\end{eqnarray} 
with
\begin{eqnarray} 
 \label{eq:delta} 
 \delta=\frac{1}{m^2}\frac{a''}{a}\ll 1. 
\end{eqnarray}

\section{Instability pockets in higher order resonances} 
\label{pocket.inst} 

We now apply the results of the previous section to the 
$\chi$ dynamics in pre-heating, by considering two cases: (i) a cubic
interaction term, $g\phi\chi^2$, between the $\phi$ and $\chi$ fields, and (ii)
the more general case of a cubic plus a quartic interaction, $g\phi\chi^2 +
h\phi^2 \chi^2$. The former case is one of the  
models that yield a Mathieu equation for the $\chi$ modes, and is here only
briefly considered  for the purpose of illustrating the usual analysis and for
comparison with our extended model (ii), 
in which the $\chi $ field equation is of the form (\ref{hill}) with $F(t)$ given by
(\ref{Ff}) and $s=2$, the simplest possible extension of the usual single frequency
Mathieu model. According to 
(\ref{widths}) and (\ref{eq:rhochi}) one would expect that, for comparable cubic and
quartic terms, the efficiency of the energy transfer process should 
approximately double in the more
general case.   

Assume  that the background field $\phi$ 
oscillates with frequency large compared with the Hubble expansion 
rate,  which is always satisfied  asymptotically  \cite{Traschen:90,Turner:1983}. 
If we neglect  the expansion in the $\phi$ dynamics then the solution of 
(\ref{eq:phi.frw}) can be given by, with a specific set of initial conditions, 
\begin{eqnarray} 
 \label{eq:phi.sol.nilH} 
 \phi(t)= A \cos (m t). 
\end{eqnarray} 

\subsection{Single frequency interaction: the Mathieu equation case}

Consider the interaction potential  $g\phi\chi^2$. This 
yields for the $k$-mode equation, using $\tau = m t$, 
\begin{eqnarray} 
\label{chitau1} 
\chi_k''+ \left[\frac{k^2}{m^2}+\frac{gA}{m^2}\cos(\tau)\right]\chi_k&=&0. 
\end{eqnarray} 
In this case one has $s=1$, and, 
\begin{eqnarray} 
 \label{eq:coefL1} 
 \omega_k^2&=&\frac{k^2}{m^2}\\ 
 c_1&=&1\\ 
 \epsilon&=& \frac{gA}{m^2}, 
\end{eqnarray} 
and thus the width of the first instability band, when it is linear on $\epsilon$ as given by (\ref{eq:Ln}), is 
\begin{eqnarray} 
 \label{eq:L1.1} 
 L_1= \frac{gA}{4m^2}. 
\end{eqnarray} 
Using (\ref{eq:rhochi}), the energy density in this case is given by
\begin{eqnarray} 
 \label{eq:rhochi1} 
 \rho_\chi^{0}(\tau)=\frac{\pi}{8} g A e^{\mu_1^{0} \tau/m}, 
\end{eqnarray} 
where $\mu_1^{0}$ is the characteristic exponent of the first band, 
evaluated at the center of the band at height $\epsilon $. 

\subsection{Multi-frequency interactions: the Hill equation case}

For the  interaction potential  $g\phi\chi^2 +h\phi^2\chi^2$ the equation for the 
$k$-th $\chi $ mode is, 
with $\tau = m t$, 
\begin{eqnarray} 
\label{chitau2} 
\chi_k''+ \left[\frac{k^2}{m^2}+\frac{hA^2}{2m^2}+\frac{gA}{m^2}\cos(\tau)+ \frac{hA^2}{2m^2}\cos(2\tau)\right]\chi_k&=&0. 
\end{eqnarray} 
In this case one has (\ref{hill}) 
and (\ref{Ff}) with $s=2$, and 
\begin{eqnarray} 
 \label{eq:coefL2} 
  \omega_k^2&=&\frac{k^2}{m^2}+\frac{hA^2}{2m^2}\\ 
 c_1&=&1\\ 
 c_2&=&\frac{hA}{2g}\\ 
 \epsilon&=& \frac{gA}{m^2}.
\end{eqnarray} 

Equations (\ref{widths}) for the widths of the first and second instability  bands, are given by (\ref{eq:Ln}) an (\ref{widths}) in the regions where they are linear in $\epsilon$, and yield
\begin{eqnarray} 
 \label{eq:L1.2} 
 L_1&=&\frac{gA}{4m^2},\\ 
 L_2&=& \frac{hA^2}{8m^2}. 
\end{eqnarray} 
The energy density in this case is given by, using (\ref{eq:rhochi}), 
\begin{eqnarray} 
 \label{eq:rhochi2} 
 \rho_\chi^{1}(\tau)= 
\frac{\pi}{8} g A e^{\mu_1^{0} \tau/m}+ 
\frac{\pi}{2} h A^2 e^{\mu_2^{0} \tau/m}, 
\end{eqnarray} 
where $\mu_1^{0}$ and $\mu_2^{0}$ are the characteristic exponents of the first 
and second bands,  evaluated at the center of the bands at height $\epsilon $.

\subsection{Reheating efficiency of the two couplings}

To proceed further in the comparison of the growth of the $\chi $ field energy
density generated by the two different couplings considered in this section, 
we must compute the relevant characteristic exponents in both cases.

In order to determine the value of the real part of the characteristic value we use the equation (\ref{eq:mue}), where the matrix $[B_{rs}]$ was truncated at 
a size of $11\times 11$ after numerical accuracy tests. 
Then we have for the value of $\alpha$ as a 
function of $\omega_k$ and of $c_2$
\begin{eqnarray} 
\alpha (c_2,\omega_k)&=& - \frac{1}{2} \left (
 \sum_{n=0}^{4} \frac{1}{(\omega_k^2-n^2)
(\omega_k^2-(n+1)^2)} \right . \nonumber \\
&& \left . + \sum_{n=0}^{3} \frac{c_2^2}{(\omega_k^2-n^2)
(\omega_k^2-(n+2)^2)} \right ) \nonumber \\
& & - \frac{c_2^2}{4 (w_k^2 - 1)^2}
\label{eq:alpha}
\end{eqnarray}
and for the value for the growth factor 
\begin{eqnarray} 
 \label{eq:mu.5} 
{\rm Re} (\mu_k)=\epsilon \alpha \sin(\pi\omega_k). 
\end{eqnarray} 
Notice that if we set $c_2=0$ in the latter expressions we recover the Mathieu case.
\begin{figure}[ht] 
\centering 
\parbox{3in}{\epsfig{file=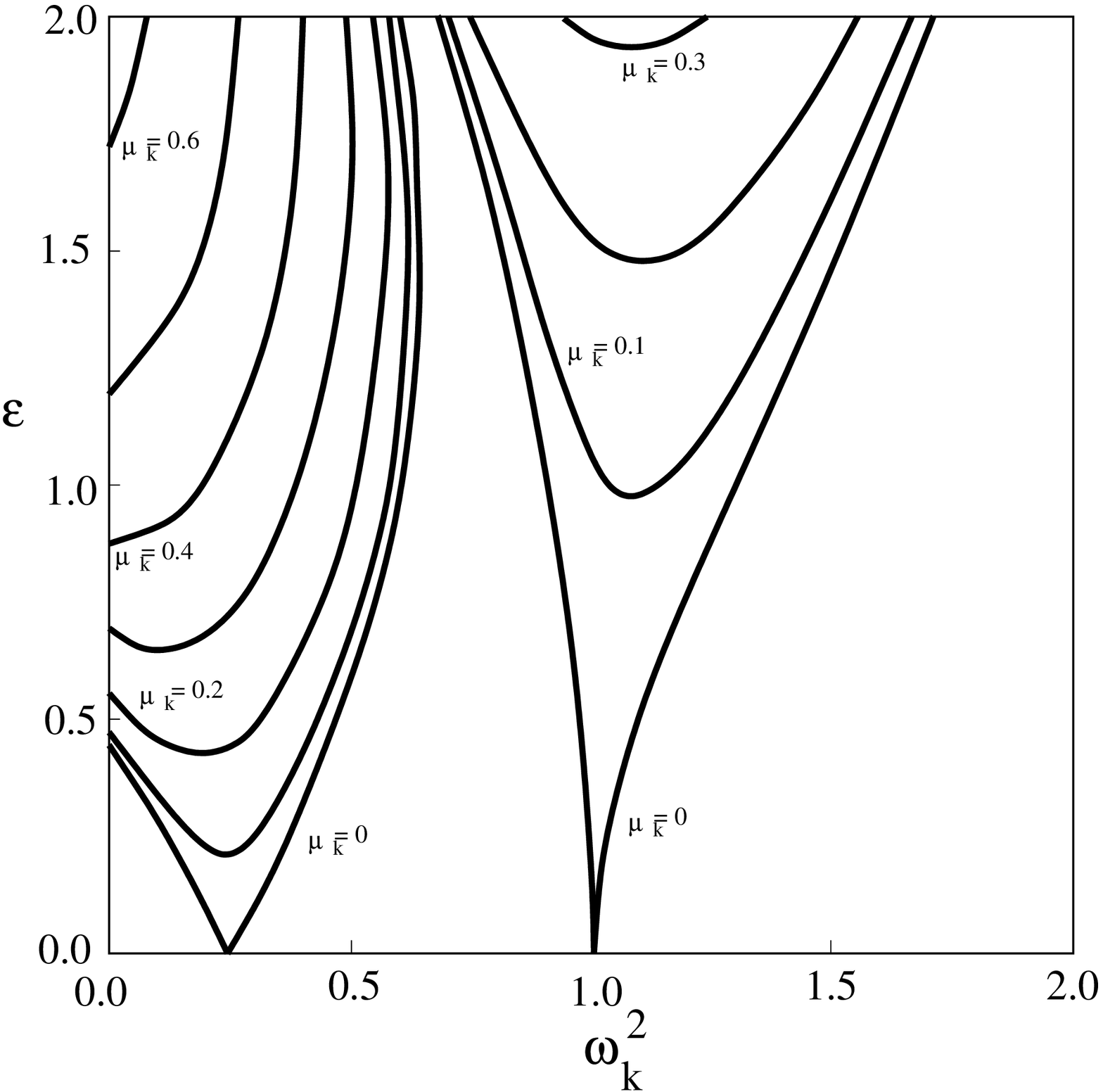, height=7cm}\figsubcap{a}} 
\parbox{3in}{\epsfig{file=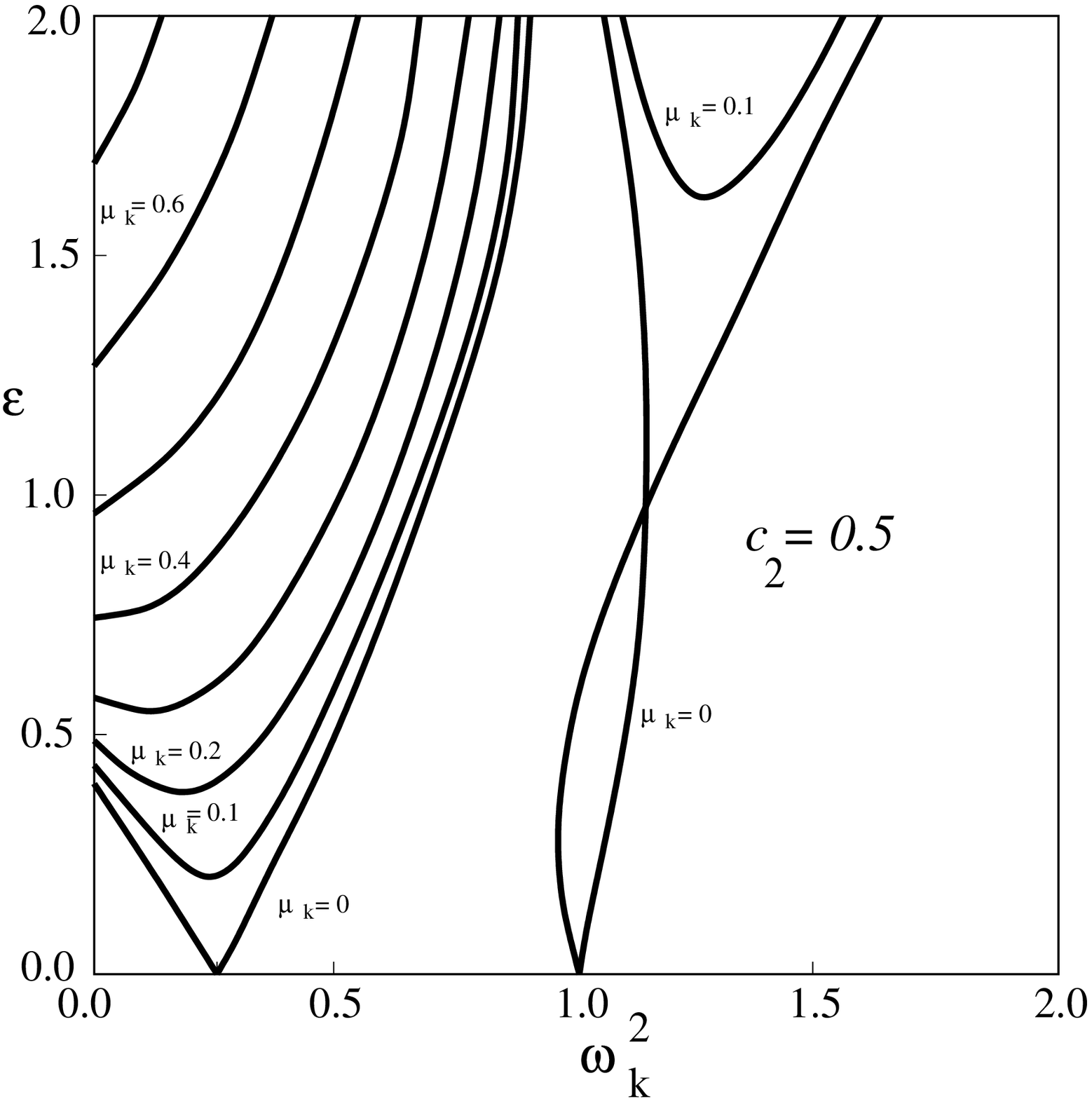, height=7cm}\figsubcap{b}} 
\caption{\footnotesize (a) Instability diagram for Mathieu equation (\ref{chitau1}); (b) Instability 
 diagram for Hill equation (\ref{chitau2}). Notice the emergence of an instability pocket in the 
 second band in Figure \ref{fig1} b), and the modifications in the level curves of the characteristic exponent $\mu$. These plots were obtained using Hill's method of solution.} 
\label{fig1} 
\end{figure} 

In Figure \ref{fig1} we show the stability diagrams for the two cases given 
above. Notice the appearance of an 'instability pocket' in the second resonance
band of equation (\ref{chitau2}) in Figure \ref{fig1} b). This phenomenon, which is much less well known than the
appearance of additional non-cuspidal instability tongues associated with
higher harmonics of the parametric forcing term, was studied in depth in 
 \cite{Broer_Levi:95, Broer1:00}. It plays a major role in explaining why
the contributions of higher order resonances may be neglected and why the $\chi $
field excitations are essentially single mode. 

As shown in Figure \ref{fig1},  there are two major differences when comparing the 
bifurcation diagram of the two-frequencies Hill case with the Mathieu case. On 
the one hand, the width of the first instability band is slightly larger and 
its level curves of constant $\mu$ are slighted tilted  when compared to the 
Mathieu case (Figure \ref{fig1}-a). On the other hand, the shape of the 
second instability  band is distorted giving rise to the emergence of a 
pocket, and its level curves of constant $\mu$ are lifted up. Also, the 
values of $\mu$ crossed by straight lines of  fixed $\epsilon$ in the first 
instability band, are significantly larger than those crossed in the second 
band, with the exception of the small region close to $\epsilon =0$.


In order to compare systematically the efficiency of the two couplings in 
transfering energy from the inflaton to the $\chi $ field, we have computed 
the time that it takes to reach an e-fold increase
of the total number of particles as a function of the coupling strength $\epsilon $, 
for $c_2=0.5$. Instead of equations (\ref{eq:rhochi1}) and (\ref{eq:rhochi2}),
which are valid only in the limit of small $\epsilon$, we use the 
general equations (\ref{eq:rhochi}) with $L_i$ taken as the numerical value of the
band widths for each model.

\begin{figure}[ht] 
\centering 
\parbox{3in}{\epsfig{file=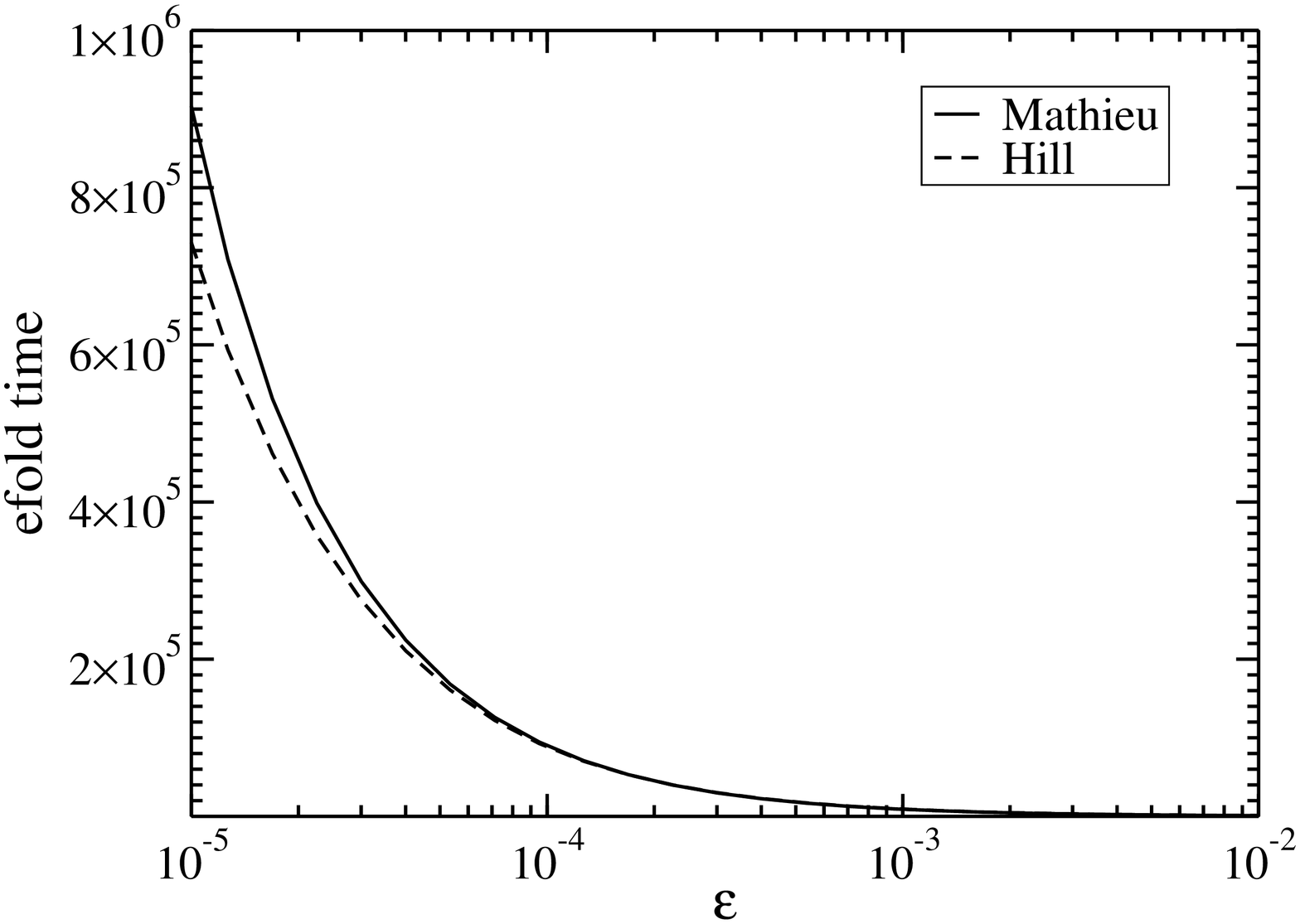, height=8cm, angle=-90}\figsubcap{a}} 
\parbox{3in}{\epsfig{file=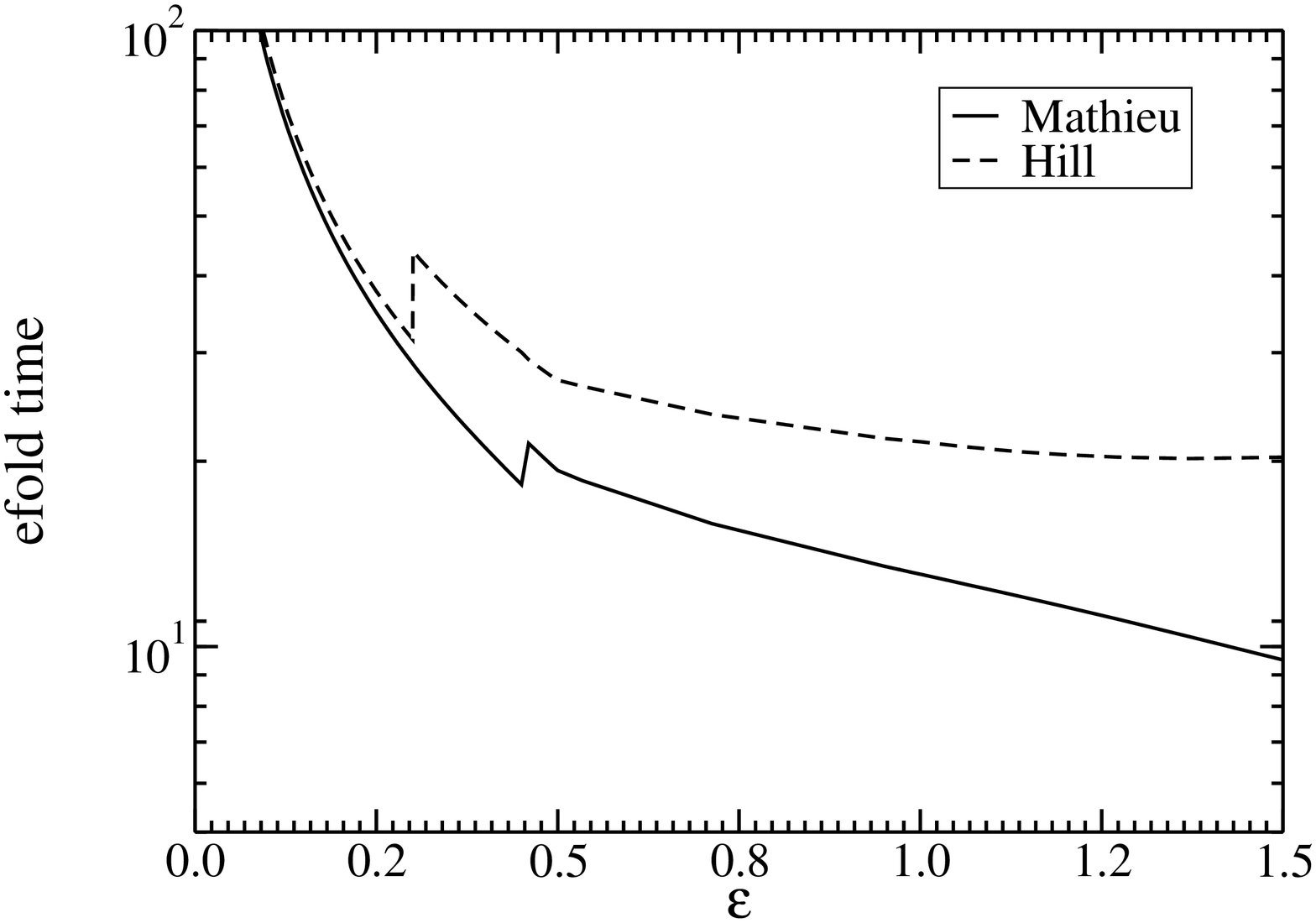, height=8cm, angle=-90}\figsubcap{b}} 
\caption{\footnotesize Time of one e-fold increase in the total number of particles
of the $\chi $ field as a function of the coupling strength $\epsilon$,
for $c_2=0.5$.} 
\label{fig2} 
\end{figure}

In Figure 2 a), we see  that the e-fold time is essentially determined by the 
contribution of the first instability band, except in the region of 
very small values of $\epsilon$. In this region, the multi-frequency coupling
becomes more efficient than the Mathieu model because of the contribution of an
additional linear instability band, but the effect has no cosmological implications
since the energy transfer achieved by both couplings is negligible for these
parameter values.

In Figure 2.b) the overall behavior of the e-fold time as a function of $\epsilon $
is shown for the two models. The kinks that can be seen in the two curves correspond 
to the values of $\epsilon $ where the first band hits the $\omega_k=0$ axis.
Due to the formation of the pocket in the
second instability band, the multi-frequency model becomes actually less
efficient than the single frequency excitation for moderate and large values
of the coupling.

This effect is also of limited cosmological relevance, since the e-fold times
of the two models are of the same order of magnitude for similar
values of the coupling strength. However, it is somehow unexpected 
that an additional linear (as opposed to cuspidal)
instability band may translate into a 
less efficient resonance mechanism for most parameter values.
This is of course a consequence of the pocket formation phenomenon,
and it shows that the conclusions based on the analytic expressions
for the asymptotic behavior of
the instability bands for small values of $\epsilon $ cannot be extrapolated.

\subsection{Reheating efficiency in the expanding universe}

We now take into consideration the expansion of the universe and discuss how this affects the contribution of the resonant bands found for the case of the two different couplings. 

The condition which should be satisfied for the resonance mechanism to work is
given by equation  (\ref{eq:cond_omega}). As discussed in section
\ref{particle.ppr} (see \cite{Traschen:90}), it translates the requirement that
the frequency of the $\chi$ particles should not be redshifted out of the
resonance  band in a time interval shorter than the amplification period
$m/(2\mu_k)$. 

Using  equation (\ref{eq:ktok}), equation (\ref{eq:cond_omega}) can be recast as
\begin{equation}
\left\vert\frac{1}{m^2}\left(\frac{a''}{a}\right)'\right\vert < |\omega_k| \left(\frac{2\mu_k}{m}\right)L_i \label{cond_expansion}
\end{equation}
for the $i$-th instability band.

According to the analysis of this section, significant particle production occurs,
for either model, only for values of $\epsilon$ such that the asymptotic approximations
(\ref{widths}) no longer hold, and that the 'instability pockets' of Figure 1.b) are
instead fully formed.
Therefore, for a given $\epsilon$, the two-frequency model will exhibit a combination of
lower values of $\mu_k$ and smaller width $L_2$ than the single frequency model.
Both effects contribute to making condition (\ref{cond_expansion}) harder to meet.
For the same coupling strength and increasing the expansion rate, the 
drift across the second instability band will become swifter than the amplification time
for the two-frequency model first.

Therefore,  when the expansion of the universe is taken into consideration, the conclusion that
the reheating mechanism is weaker in the multi-frequency case than in the  
Mathieu case is  reinforced. Since, however, only the contribution of the first instability
band is accounted for in the efficiency estimates found in the literature, 
these estimates remain valid for the multi-frequency model.

Now equations (\ref{eq:rhochi1}) and (\ref{eq:rhochi2}) are valid as long as any 
given mode remains in the resonance band. Due to the expansion, the time interval $\Delta \tau$ during which a mode remains in the 
band is 
\begin{eqnarray} 
 \label{eq:dtau} 
 \Delta\tau \simeq \frac{L}{H m^2}. 
\end{eqnarray} 
As long as the total time is small compared with $H^{-1}/m$, the total energy produced during the time 
interval $N\Delta \tau$ is approximately given by $N\rho_\chi$. 
In this scenario, reheating is efficient if the ratio 
\begin{eqnarray} 
 \label{eq:ratio} 
 \frac{N\rho_\chi}{\rho_\phi} 
\end{eqnarray} 
becomes of order one after a time smaller than the Hubble time. Otherwise  a 
significant fraction of the original energy density is redshifted away. 

For the first band (as we have seen, the only one that might contribute), the latter quantity (\ref{eq:ratio}) reaches the value one for 
\begin{eqnarray} 
 \label{eq:N1} 
 N=\frac{\rho_\phi}{\rho_\chi}\simeq\frac{m^2A^2}{\frac{\pi}{8}gA e^{\mu \Delta \tau/m}}. 
\end{eqnarray} 
Therefore the condition for sufficient reheating $N\Delta\tau<H^{-1}/m$ becomes 
\begin{eqnarray} 
 \label{eq:N1.1} 
 \mu\frac{\Delta\tau}{m} e^{\mu \Delta \tau/m} >\frac{24}{H^3\alpha}{\pi m^2}, 
\end{eqnarray} 
or, 
\begin{eqnarray} 
 \label{eq:N1.1.aix} 
 \mu\frac{\Delta\tau}{m}> W\left(\frac{24}{H^3\alpha}{\pi m^2}\right), 
\end{eqnarray} 
where $W$ is Lambert function \cite{Lambert 96}. Equations (\ref{eq:dtau}) and
 (\ref{eq:N1.1.aix}) establish a relation between the free parameters for
 sufficient reheating to take place that holds for both models considered in
 this section.  

\section{Conclusions} 
We have shown how the consideration of extra frequencies in the $\chi$ equation
of motion can be found in a simple model with  
$V(\phi)=m^2\phi^2 /2 $ for the inflaton potential
and the interaction potential 
$V(\phi,\chi)=g\phi\chi^2+h\phi^2 \chi^2$. In the narrow resonance regime the
phenomenon of parametrically 
resonant excitations of the scalar field $\chi$ by the inflaton's oscillations
is then governed by a  Hill equation, where  
the forcing term in the $\chi$ equation has two distinct and commensurable frequencies. 

As a result of the presence of various harmonics in the equation of 
motion of the $\chi$ field, the geometrical features of the resonant bands in
the bifurcation parameter space are modified. Two main changes take place with
respect 
to the single frequency (Mathieu) case. On the one hand, for small amplitudes
of excitation, 
there are two resonant tongues with linear dependences on the amplitude, rather
than just one as in the Mathieu case. On the other hand, closely related to the
previous effect, there is a distortion of this additional band which gives rise
to the formation of 'instability 
pockets'. Due to the simultaneous presence (and interference) of two excitation frequencies,
the lines in parameter space that correspond to the periodic solutions ($\mu=0$) and define
the boundary of the instability band cross each other for values of the forcing
amplitude of order one.

For the two cases under consideration we have evaluated and compared the 
particle production rates.
We have considered first that the  inflaton field $\phi$ oscillates around the
minimum of its potential much faster than the expansion rate of the universe,
$m_\phi\gg H$, so that the expansion of the universe may be neglected. 
We have shown that, in general, the presence of an additional excitation frequency hinders,
rather than favors, the efficiency of parametric resonance as an energy transfer mechanism,
and that this is a consequence of the 'instability pockets' in the bifurcation diagram
of the general Hill's equation.
The enhancement of particle production due to the presence of a second linear, as opposed to
cuspidal, instability band is shown to occur only for extremely small coupling strengths, 
for which both models yield negligible rates of $\chi $ particle creation.
We then argue that the effects of an expanding universe further justify 
the approximation of
neglecting the contribution of the second non-cuspidal instability band.

In conclusion, our detailed analysis of two different coupling terms supports
and justifies the usual approach in the literature, where the efficiency of
reheating by parametric resonance 
is evaluated by considering the simplest form of the parametrically forced
equation, and only the dominant contribution of the first instability band. 

\section*{Acknowledgements}

Financial support from the Foundation of the University of Lisbon 
and the Portuguese Foundation for Science and 
Technology (FCT) under contracts POCI/FP/ FNU/50216/2003
and POCTI/ISFL/2/618 is gratefully 
acknowledged.

 

\begin{thebibliography}{02} 
\bibitem{Shtanov:95} Y. Shtanov, J. Traschen and B. Brandenberger 
Phys. Rev. D {\bf 51}, 5438 (1995) 

\bibitem{Kofman:1997b}
L.~Kofman, A.~D.~Linde and A.~A.~Starobinsky,
Phys.\ Rev.\ D {\bf 56} (1997) 3258
[arXiv:hep-ph/9704452].

\bibitem{Berera 95a}
A. Berera, Phys. Rev. Lett., 75:3218 (1995).

\bibitem{Mimoso:2005bv}
  J.~P.~Mimoso, A.~Nunes and D.~Pavon,
  Phys.\ Rev.\  D {\bf 73}, 023502 (2006)
  [arXiv:gr-qc/0512057].

\bibitem{Liddle+Lyth:2000} A. R. Liddle and D. H. Lyth, 
{\em Cosmological Inflation and Large-Scale-Structure}  (Cambridge University Press, cambridge, England, 2000).

\bibitem{Kofman:1994rk}
L.~Kofman, A.~D.~Linde and A.~A.~Starobinsky,
Phys.\ Rev.\ Lett.\  {\bf 73} (1994) 3195
[arXiv:hep-th/9405187].

\bibitem{Boyanovsky+1:95} D. Boyanovski, H. J. de Vega, R. Holman, 
 D. S. Lee and A. Singh 
Phys. Rev. D {\bf 51}, 4419 (1995) 

\bibitem{Boyanovsky+2:95} D. Boyanovski, M. D'Attanasio, H. J. de 
 Vega, R. Holman, D. S. Lee and A. Singh 
Phys. Rev. D {\bf 52}, 6805 (1995) 

\bibitem{Yoshimura:1995gc}
M.~Yoshimura,
Prog.\ Theor.\ Phys.\  {\bf 94}, 873 (1995)
[arXiv:hep-th/9506176].

\bibitem{Fujisaki:1995dy}
H.~Fujisaki, K.~Kumekawa, M.~Yamaguchi and M.~Yoshimura,
Phys.\ Rev.\ D {\bf 53}, 6805 (1996)
[arXiv:hep-ph/9508378].

\bibitem{Chung:1998rq}
D.~J.~Chung, E.~W.~Kolb and A.~Riotto,
Phys.\ Rev.\ D {\bf 60} (1999) 063504
[arXiv:hep-ph/9809453].

\bibitem{Berges:2002cz}
  J.~Berges and J.~Serreau,
  Phys.\ Rev.\ Lett.\  {\bf 91}, 111601 (2003)
  [arXiv:hep-ph/0208070].

\bibitem{Dolgov:1989us}
  A.~D.~Dolgov and D.~P.~Kirilova,
  Sov.\ J.\ Nucl.\ Phys.\  {\bf 51}, 172 (1990)
  [Yad.\ Fiz.\  {\bf 51}, 273 (1990)].

\bibitem{Traschen:90} 
J.~H.~Traschen and R.~H.~Brandenberger,  
Phys. Rev. D, {\textbf{42}}, 8, 2491--2504, (1990).

\bibitem{Kaiser:1997hg}
D.~I.~Kaiser,
Phys.\ Rev.\ D {\bf 57} (1998) 702
[arXiv:hep-ph/9707516].

\bibitem{Greene:1997fu}
P.~B.~Greene, L.~Kofman, A.~D.~Linde and A.~A.~Starobinsky,
Phys.\ Rev.\ D {\bf 56} (1997) 6175
[arXiv:hep-ph/9705347].

\bibitem{Bassett:2005xm}
  B.~A.~Bassett, S.~Tsujikawa and D.~Wands,
  Rev.\ Mod.\ Phys.\  {\bf 78}, 537 (2006)
  [arXiv:astro-ph/0507632].


\bibitem{Basset:98} B. A. Basset 
Phys. Rev. D {\bf 58} (1998) 021303 
[arXiv:hep-ph/9709443v3]. 

\bibitem{Bassett:1999}
B.~A.~Bassett, F.~Tamburini, D.~I.~Kaiser and R.~Maartens,
Nucl.\ Phys.\ B {\bf 561} (1999) 188
[arXiv:hep-ph/9901319].



\bibitem{Anderson:1996kr}
  G.~W.~Anderson, A.~D.~Linde and A.~Riotto,
  Phys.\ Rev.\ Lett.\  {\bf 77}, 3716 (1996)
  [arXiv:hep-ph/9606416].

\bibitem{Son:1996uv}
  D.~T.~Son,
  Phys.\ Rev.\  D {\bf 54}, 3745 (1996)
  [arXiv:hep-ph/9604340].


\bibitem{Kasuya:1997ha}
  S.~Kasuya and M.~Kawasaki,
  Phys.\ Rev.\  D {\bf 56}, 7597 (1997)
  [arXiv:hep-ph/9703354].

\bibitem{Khlebnikov:1998sz}
  S.~Khlebnikov, L.~Kofman, A.~D.~Linde and I.~Tkachev,
  Phys.\ Rev.\ Lett.\  {\bf 81} (1998) 2012
  [arXiv:hep-ph/9804425].

\bibitem{Micha:2002ey}
  R.~Micha and I.~I.~Tkachev,
  Phys.\ Rev.\ Lett.\  {\bf 90}, 121301 (2003)
  [arXiv:hep-ph/0210202].
\bibitem{Micha:2003ws}
  R.~Micha and I.~I.~Tkachev,
  arXiv:hep-ph/0301249.

\bibitem{Podolsky:2005bw}
  D.~I.~Podolsky, G.~N.~Felder, L.~Kofman and M.~Peloso,
  Phys.\ Rev.\  D {\bf 73}, 023501 (2006)
  [arXiv:hep-ph/0507096].


\bibitem{Charters:2005eg}
  T.~Charters, A.~Nunes and J.~P.~Mimoso,
  Phys.\ Rev.\  D {\bf 71} (2005) 083515
  [arXiv:hep-ph/0502053].

\bibitem{Felder:2006cc}
  G.~N.~Felder and L.~Kofman,
  Phys.\ Rev.\  D {\bf 75}, 043518 (2007)
  [arXiv:hep-ph/0606256].
\bibitem{GarciaBellido:2007dg}
  J.~Garcia-Bellido and D.~G.~Figueroa,
  Phys.\ Rev.\ Lett.\  {\bf 98}, 061302 (2007)
  [arXiv:astro-ph/0701014].

\bibitem{Finelli:1998bu}
  F.~Finelli and R.~H.~Brandenberger,
  Phys.\ Rev.\ Lett.\  {\bf 82}, 1362 (1999)
  [arXiv:hep-ph/9809490].


\bibitem{Tsujikawa:2002nf}
  S.~Tsujikawa and B.~A.~Bassett,
  Phys.\ Lett.\  B {\bf 536}, 9 (2002)
  [arXiv:astro-ph/0204031].

\bibitem{Jokinen:2005by}
  A.~Jokinen and A.~Mazumdar,
  JCAP {\bf 0604}, 003 (2006)
  [arXiv:astro-ph/0512368].

\bibitem{Sa:2007pc}
  P.~M.~Sa and A.~B.~Henriques,
  Phys.\ Rev.\  D {\bf 77}, 064002 (2008)
  [arXiv:0712.2697 [astro-ph]].

\bibitem{Felder:2000hj}
  G.~N.~Felder, J.~Garcia-Bellido, P.~B.~Greene, L.~Kofman, A.~D.~Linde and I.~Tkachev,
  Phys.\ Rev.\ Lett.\  {\bf 87}, 011601 (2001)
  [arXiv:hep-ph/0012142].


\bibitem{Taruya:1997iv}
  A.~Taruya and Y.~Nambu,
  Phys.\ Lett.\  B {\bf 428}, 37 (1998)
  [arXiv:gr-qc/9709035].

\bibitem{Desroche:2005yt}
  M.~Desroche, G.~N.~Felder, J.~M.~Kratochvil and A.~Linde,
  Phys.\ Rev.\  D {\bf 71}, 103516 (2005)
  [arXiv:hep-th/0501080].



\bibitem{Dufaux:2006ee}
  J.~F.~Dufaux, G.~N.~Felder, L.~Kofman, M.~Peloso and D.~Podolsky,
  JCAP {\bf 0607}, 006 (2006)
  [arXiv:hep-ph/0602144].




\bibitem{BasteroGil:2007mm}
  M.~Bastero-Gil, M.~Tristram, J.~F.~Macias-Perez and D.~Santos,
  Phys.\ Rev.\  D {\bf 77}, 023520 (2008)
  [arXiv:0709.3510 [astro-ph]].

\bibitem{ArmendarizPicon:2007iv}
  C.~Armendariz-Picon, M.~Trodden and E.~J.~West,
  JCAP {\bf 0804}, 036 (2008)
  [arXiv:0707.2177 [hep-ph]].

\bibitem{Arnold} V. Arnold,  {\em Ordinary Differential Equations} The MIT Press (1978)

\bibitem{Broer_Levi:95} H. Broer and M. Levi 
Arch. Rational Mechanics Anl. 131, 225-240, 1995. 

\bibitem{Broer1:00} H. Broer and C. Sim\'o 
J. Dif. Eq., 166 (2000), 290-327. 



\bibitem{Turner:1983} M. S. Turner,
Phys.\ Rev.\ D {\bf 28} (1983) 1243.


\bibitem{HillsEqBook} 
W. Magnus,  and S. Winkler, 
``Hill's equation'', Dover, (1979) 

\bibitem{Lambert 96}
Robert M. Corless, G. H. Gonnet, D. E. G. Hare, D. J. Jeffrey, and D. E. Knuth, 
Advances in Computational Mathematics, 5 (1996) 329--359. 

\end{thebibliography}
\end{document}